\documentclass[aps,twocolumn,pre,showpacs]{revtex4-1}
\usepackage{amsmath}
\usepackage{graphicx}
\usepackage{bm}
\usepackage{amssymb}
\usepackage{array}

\graphicspath{{figures/}}

\newcommand{\dd}[0]{\text{d}}
\newcommand{\CG}[0]{C_{\text{G}}}
\newcommand{\CQ}[0]{C_{\text{Q}}}
\newcommand{\CinsOne}[0]{C_{\text{ins},1}}
\newcommand{\CinsTwo}[0]{C_{\text{ins},2}}
\newcommand{\Cext}[0]{C_{\text{ext}}}
\newcommand{\Cgeo}[0]{C_{\text{geo}}}
\newcommand{\ncrit}[0]{n_\text{crit}}

\begin{document}
\title{Mechanical Strain can Switch the Sign of Quantum Capacitance from Positive to Negative}
\author{Yuranan Hanlumyuang$^1$, Xiaobao Li$^{2}$, Pradeep Sharma$^{2,3}$$^\clubsuit$ \\
{\em $^1$ Department of Materials Engineering, Faculty of Engineering, Kasetsart University, Bangkok, 10900, Thailand}\\
{\em $^2$ Department of Mechanical Engineering, University of Houston, TX, 77204, USA}\\
{\em $^3$ Department of Physics, University of Houston, TX, 77204, USA}
}
\date{\today}


\begin{abstract}
Quantum capacitance is a fundamental quantity that can directly reveal many-body interactions among electrons 
and is expected to play a critical role in nanoelectronics. One of many tantalizing recent physical revelations 
about quantum capacitance is that it can posses a \emph{negative} value, hence  allowing for the possibility of \emph{enhancing} 
the overall capacitance in some particular material systems beyond the scaling predicted by classical electrostatics. 
Using detailed quantum mechanical simulations, we find an intriguing result that mechanical strains can tune both \emph{signs} 
and \emph{values} of quantum capacitance.
We use a small coaxially-gated carbon nanotube as a paradigmatical capacitor system and show that, for the range of mechanical strain considered, quantum capacitance 
can be adjusted from very large positive to very large negative values (in the order of plus/minus hundreds of attofarads), 
compared to the corresponding classical geometric value (0.31035  aF).
We elucidate the mechanisms underpinning the switching of the sign of quantum capacitance due to strain. 
This finding opens novel avenues in designing quantum capacitance for 
applications in nanosensors, energy storage, and nanoelectronics.
\end{abstract}

\maketitle

The behavior of nanoscale capacitors is remarkably rich and exhibits features unanticipated by conventional electrostatic theory \cite{Buttiker93, Spaldin06,  Spaldin09}.
Intense research activity has recently ensued on materials development, elucidation of the fundamental science and applications 
related to nano capacitors. Conventional wisdom, with its origins in text-book electrostatics, suggests that high capacitance can 
be achieved by reducing the characteristic size of the dielectric materials with high dielectric permittivity. For example, in the 
case of a thin film configuration, the classical electrostatic capacitance per unit area is taken to be $\Cgeo=\epsilon/d$ where $\epsilon$ 
is the dielectric permittivity and $d$ is the thickness of the dielectric sandwiched between the electrodes. Accordingly,  materials development 
have tended to focus on the selection and engineering of high dielectric permittivity materials at the nanoscale along with the concomitant 
challenges of their fabrication and testing. Unfortunately, experiments have shown that in many thin and high-permittivity insulators
(such as Ta$_2$O$_5$), their capacitance values are orders of magnitude smaller \cite{Mead61, Saad04, Boesch08}---compared to 
what may be expected from classical electrostatics. Using \emph{ab inito} calculations and many-body quantum theory, 
this perplexing discovery has been explained in part by the existence of the atomically thin two-dimensional electron gas (2DEG) between
 the insulators and the metallic gates \cite{Spaldin06, Spaldin09, Kopp09}. 
The electron accumulation at such interfaces is due to the imperfect screening at the metallic interface, causing an extraneous 
contribution to overall capacitance,  termed \emph{quantum capacitance}. In the metallic (gate)-insulator-metallic (gate) setup,  
the quantum capacitance is incorporated into a series capacitor model as:
\begin{equation}
\frac{1}{\CG} = \frac{1}{\Cgeo} + \frac{2}{\CQ}.
\label{eqn:1}
\end{equation}
\noindent
Here $\CG$ is the capacitance measured by mounting  leads to an external voltmeter, $\Cgeo$ is the classical geometric capacitance 
obtained from electrostatic principles, and $\CQ$ is the aforementioned quantum capacitance.  In general $\CQ$ is many orders of 
magnitude greater than the geometric capacitance $\Cgeo$, hence its effects in sub-micron or larger scales are nearly nonexistent.  $C_Q$ can be only 
realized in nanoscale or in devices containing high-dielectric materials.

Despite having been identified for nearly three decades \cite{Luryi88}, quantum capacitance has only gained much interest for 
in recent years \cite{Ilani06, Li11,Liu12,Yu13}.  From first-principles viewpoint, quantum capacitance arises 
from many-body interactions among electrons, and it relates to the change in electron density with the chemical potential of the 2DEG as 
\begin{equation}
\frac{1}{\CQ} = \frac{\dd\mu/\dd n}{ A e^2}.
\label{eqn:2}
\end{equation}
Contributions to $\dd\mu/\dd n$ stems from density of states, and exchange/correlation energy functional of the 2DEG.  At a glance, 
it may seem as if quantum capacitance will always tend to diminish the overall capacitance $\CG$.  However, due to the negative 
exchange-correlation contribution, quantum capacitance can very well enhance the overall capacitance.  Recently, a few 
materials systems  have been identified in experiments as plausible realization of this enhancement effects. Examples 
include C-CuO$_2$-Cu coaxial nanowires and LaAlO$_{3}$/SrTiO$_3$ films system where about $100$\% and 40\% 
enhancement of capacitance have been reported respectively \cite{Li11, Liu12}.

A number of quantitative effects, including the finite thickness of the two-dimensional sheets, interface reconstruction, 
among many others can influence quantum capacitance \cite{Kopp09}.  We report here a surprising aspect of the 
quantum capacitance pertaining to mechanical deformation. This notion may be motivated by considering a 2DEG 
in low-electronic density limit. A dimensionless  distance $r_s$ may be defined as $r_s \sim n^{-1/2}$ where $n$ is 
the total carrier density. 
This distance characterizes the inter particle distances which directly depend on the amount of applied strain.  
Using this argument, we rationalize that mechanical strain should alter the electronic densities, and the exchange 
contribution to quantum capacitance. Although  existing studies have already established the fundamental science 
underlying quantum capacitance \cite{Giuliani05, Fogler05, Kopp09, Kusminskiy08, Borghi10}, the potential role of mechanical effects  remains  unexplored. 

In this Letter, we explore the possibility of tuning quantum capacitance (hence the overall capacitance $\CG$) via mechanical strain.  We find that, for a model system based on carbon nanotube,  in conjunction with appropriate doping levels, mechanical strain can indeed substantially change the values of quantum capacitance and more intriguingly, switch its sign from positive to negative (and vice versa).

\begin{figure}
\centering
$\begin{array}{c}
\includegraphics[clip,scale=0.4]{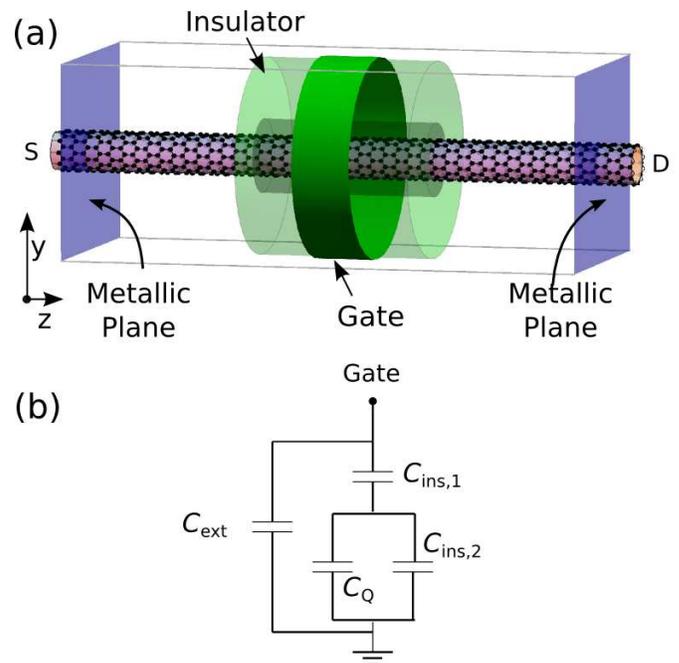} \\
\end{array}
$
\caption{\label{fig:simBox} (a) In a geometric setup for a DFTB+NEGF calculation,  
a carbon nanotube is located inside the Poisson box.  
The source and drain of electrons are marked by the letters S and D respectively. 
(b) Circuital model of the cylindrically-gated carbon nanotube. }
\end{figure}

The model configuration is shown in Fig. \ref{fig:simBox}(a).  A (10,0) nanotube is coaxially gated
by a metallic stripe. The stripe is separated from the nanotube by a SiO$_2$ insulator with the relative permittivity of $\epsilon = 3.9$.  All
features are placed inside a Poisson box where the electronic charge and potential is solved self-consistently. Two metallic planes
act as a grounded reference contacts which serve as source/drain of electrons in response to applied fields of the cylindrical gate.

Quantum capacitance has been obtained using a combination of Density Functional Tight-Binding Theory (DFTB) and 
methods in nonequilibrium Green's functions (NEGF)---the details related to the specific computational code may be found in Refs. \cite{Aradi07, Pecchia04, DiCarlo05}. The exchange-correlation (XC) contribution to the total electronic energy is  approximated by the Hubbard model. The carbon-carbon interactions are those provided in the Slater-Koster parameter set in \cite{Latessa05}, and PBC-0-1 \cite{Rauls99}. 
The DFTB+NEGF package used \cite{Aradi07} has been modified to allow for the aforementioned coaxially gated nanotube configuration.  
The underlying principles  of the setup will become clear in a moment. Emphasizing first on model description, the nanotube considered 
contains 2,160 atoms and has about 22.6 nm in length.  It has been coaxially  gated by a  hollow metallic cylindrical stripe located at a distance about 11.1 \AA$\,$ away.
The separating insulator gives rise to the regular geometric capacitance ($\Cgeo$) as expected.  Dirichlet boundary conditions have been enforced on the metallic planes to simulate grounded contacts at source and drain. 
The same boundary condition is also used to apply a constant voltage at the cylindrical metallic gate.  The source/drain outside of the Poisson box are used for induced charge accumulation (or depletion) at the surface of the nanotube which is treated within the NEGF framework.  Our geometric set up follows ones in  \cite{Latessa05, Latessa07}. 

It is well accepted that the exchange contribution to the total energy of a solid depends strongly on the number of charge carriers \cite{Kopp09}. Here, the number of charge carriers  is allowed to vary continuously by fractional n-doping of the nanotube.  Doping is simulated by adjusting numbers of fractional electron in valence orbital of each carbon atom.  In this way, the number of electron filling in the conduction subband can be tuned continuously. This numerical doping routine has been proved successful in studying capacitance of various mechanically-neutral one-dimensional structures \cite{Latessa05, Latessa07}.  Ultimately, the induced charge on the cylindrical surface enables us to numerically extract quantum capacitance.

\begin{figure}
\includegraphics[clip,scale=0.6]{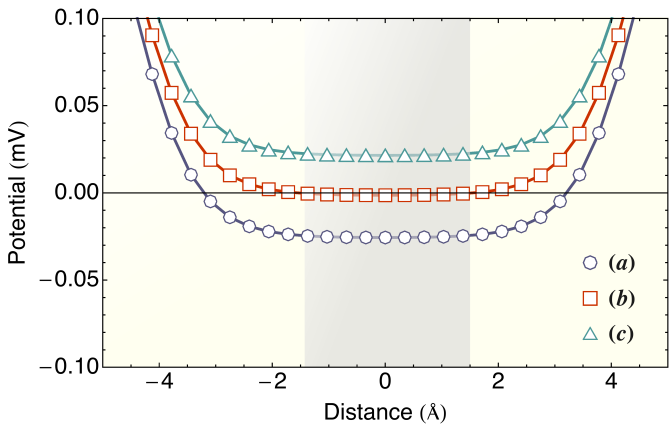}
\caption{\label{fig:voltages} Electrostatic potentials along the $y-$direction of a mechanically neutral (10,0) nanotube. 
The tube is outlined in the central shaded area.  
Three screening regimes, depending upon carrier densities, are illustrated. 
(a) For $n = 0.1033$ \AA$^{-1}$, the CNT overcompensate the gate field and accumulates more electrons
than needed, leading to \emph{negative} potential in the interior of the tube, i.e.  the nanotube \emph{overscreen} the 
applied electric fields.  (b) At a $n = 0.1005 $ \AA$^{-1}$, the accumulated
charge on the CNT completely screens the gate field. Zero screened potential  is present at the center of the CNT interior. 
(c) At the carrier density $n =  0.0478 $ \AA $^{-1}$, the charge accumulation
on the CNT is limited by the density of states  \cite{Luryi88}, 
and consequently the CNT only partially screens the positive gate field. A small positive potential is then present in
the interior of the CNT.
}
\end{figure}

The model of all-around gated nanotube, inspired by the seminal paper of Luryi \cite{Luryi88}, can be simplified 
into an equivalent circuital model shown in Fig. \ref{fig:simBox}(b).  A modification beyond the original work is an 
addition external capacitance ($C_{\text{ext}}$) in parallel to the main circuit. This external capacitance accounts for possible charge accumulations 
at the metallic planes outer to the nanotube.  The capacitance $\CinsOne$ accounts for the separating oxide insulator, while $\CinsTwo$ 
quantifies the accumulated charge at the metallic plates inner to the CNT.  The induced charge on the surface of the nanotube leads to 
quantum capacitance, denoted by $\CQ$. The numerical values of $\CQ(n)$ as a function of carrier density
can be straightforwardly obtained after establishing relations between electrostatic potential profiles and doping level $n$. 
Fig.  \ref{fig:voltages} shows  three regimes of potential profiles along the $y-$direction (shown in Fig. \ref{fig:simBox}) 
where the carrier densities are (a) $n = 0.1033$, (b) $0.1005$,  and (c) $0.0478$  \AA$^{-1}$ respectively, and the applied gate voltage is $\delta V_\text{G} = 1$ mV.
Explanations of these potential regimes in relations to carrier densities are detailed in the caption. The nanotube either (a) overscreens, 
(b) completely screens, or (c) partially screens the gate potential depending on the carrier densities \cite{Giuliani05}. This screening 
behavior of the nanotube are largely caused by the exchange energy function (through varying carrier densities). 

The preceding paragraphs and the accompanying figures make it clear that the rather specific geometric setup of the 
nanotube, insulator, metallic gate and planes allows for direct calculations of $\CQ$.  Taking advantage of 
the three screening regimes and the circuit model shown in Fig. \ref{fig:simBox}(b), the quantum capacitance $\CQ(n)$ is obtained
by computing the number of induced charge as a function of the carrier densities. 
By numerically evaluating the  charged induced on the metallic planes shown in Fig. \ref{fig:simBox} (a), we find that 
both $\CinsTwo$ and $\Cext$ for all current long CNTs studied are negligible. The gate capacitance in the circuit model in Fig. \ref{fig:simBox}(b)
hence reduces to $1/\CG = 1/\CQ+ 1/\CinsOne$. In the case of complete screening, we have 
$1/\CG =  1/\CinsOne  = \delta V_\text{G}/\delta Q(\ncrit) $
whereas, for partial screening or over screening
$1/\CG' = 1/\CinsOne
+1/\CQ(n) =\delta V_\text{G}/\delta Q(n)$.
Since the gate voltage is externally controlled, 
$\delta V_\text{G}$ in both cases can be equated to yield the quantum capacitance as
\begin{equation}
\CQ(n) = \CinsOne\left[\frac{\delta Q(\ncrit)}{\delta Q(n)}-1\right]^{-1},
\label{eqn:5}
\end{equation}
\noindent
where $\ncrit$ is a critical value of the carrier densities when complete screening is observed. The task of computing quantum 
capacitance then reduces to determining $\ncrit$, and the induced charge at each doping levels.  Here, an additional numerical optimizing routine 
has been implemented in coordination with the DFTB+NEGF package \cite{Aradi07, Pecchia04, DiCarlo05} in order to obtain the induced charge. 
This numerical routine appropriately sets Fermi levels, necessary for integrating the 
energy subbands for charge densities.   A necessary condition to determine the Fermi levels is the charge neutrality. 
The charge tolerance in the Fermi energy optimizing routine 
is set to  $\delta Q  = 10^{-7} $ a.u. After obtaining appropriate Fermi energies,  the NEGF routine is then utilized to draw electrons from
source and drain onto the cylindrical surface of the nanotube \cite{Pecchia04, DiCarlo05}.


\begin{figure}
\centering
$ 
\begin{array}{c}
\includegraphics[clip, scale=0.6]{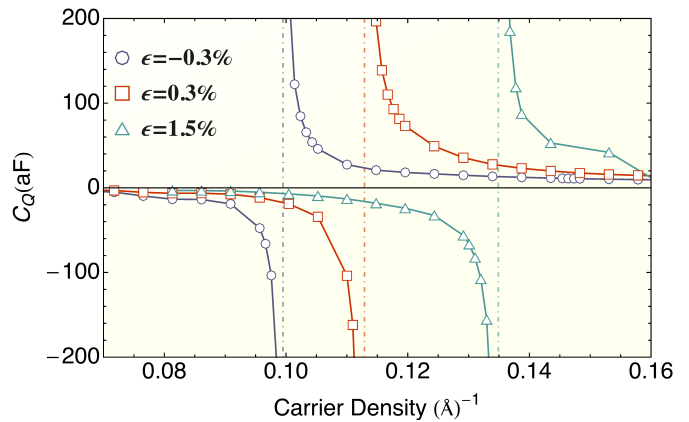} 
\end{array}
$
\caption{\label{fig:strain} The effect of strain on quantum capacitance.}
\end{figure}

Small uniaxial stretch and compression can be applied the nanotube using the relation $z\mapsto (1+\epsilon) z$.  
Quantum capacitances at three strain levels are shown in Fig. \ref{fig:strain}.  As well-evident from the figure, for 
an appropriate doping level, $\CQ$ can change both its sign and magnitude depending upon the level of applied mechanical strain. 
For example at the carrier density of about 0.10  \AA$^{-1}$, $\CQ$ changes from very high positive value at the $\epsilon = -0.3 $\% to 
about $\CQ \approx -5 $ aF at $\epsilon = 1.5$ \%. For the (10,0) nanotube considered, this change in $\CQ$ amounts
for about $10$ \% change in the overall capacitance $\CG$ in the classical complete screening case. Computed
from from expressions leading Eq.(\ref{eqn:5}), the classical capacitance is $\CinsOne =  0.31035 $ aF. 
The change in sign at the carrier density $n = 0.10$ \AA$^{-1}$ can be interpreted as follows. 

\begin{figure}
\includegraphics[clip,scale=0.3]{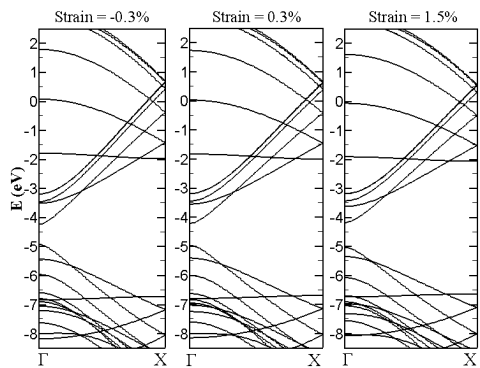}
\caption{\label{fig:band} Dependence of CNT band diagrams on mechanical strain.}
\end{figure}

As the level of strains goes from small compression to tensile, the electronic density decreases due to an increase in the cylindrical tube area.  
At a sufficiently low electron density, the exchange contribution $E_x[n]$ dominates the kinetic part to the capacitance. The exchange energy alone overwhelms and 
produces a negative quantum capacitance.  By examining the band diagrams of the electronic ground state of the nanotube in Fig. \ref{fig:band} at different strain levels, it is clear that  densities of states are not altered much across the range of strain considered. The kinetic contribution, which manifests through the density of states \cite{Kopp09}, 
thus plays a less crucial role in tuning magnitude and sign of $\CQ$ in the nanotube under study. 

In summary,  we have explored the notion of switching the sign of quantum capacitance via mechanical strain, using a coaxially gated carbon nanotube as a model material system. From full electronic structure calculations within density functional tight-binding  theory,  it is clear that the interplay between doping and exchange energy functional play crucial roles in determining both the sign and magnitude of quantum capacitance.  Fig. \ref{fig:strain} clearly demonstrates that the quantum capacitance can be mechanically controlled from very large positive to very large negative values.  It is then natural to ask if this nanotube property can be utilized in nanosensing technology. We envision that one possible technique to detect such a large negative/positive quantum capacitance is to modify the ingenious apparatus in \cite{Eisenstein92}. A pair of graphene could be used as place holders for 2DEG. The graphene could be set on piezoelectric materials to simulate strains, while small gates can be intentionally attached to add/remove electrons to the system. Strong signals of ratio between differential change in gate electric field and the penetrating fields ought to be observed.  As discussed in \cite{Eisenstein92},  the ratio correlates directly to quantum capacitance. It should be emphasized that the concept of quantum capacitance is not only limited to CNT.  Exploring the mechanical
effects in other material systems such as BN monolayers,  LaAlO$_3$/SrTiO$_3$ films could very well open new possibility 
in improving lower-power devices, and in energy storage for nanoelectronics.  Moreover, the switching of sign promises a 
novel ``quantum sensing" mechanism.

\begin{acknowledgements}
YH and  XL acknowledge useful correspondences with Alessandro Pecchia,  Aldo Di Carlo, B\'alint Aradi
and  Gabriele Penazzi during the course of this work. 
\end{acknowledgements}


\end{document}